\documentclass[letterpaper,11pt]{article}
\usepackage{amsmath,mathrsfs,amssymb,amsfonts,amsthm,mathtools}
\usepackage{physics,color}
\usepackage{graphicx}
\usepackage{subfigure}
\usepackage{tabularx} 
\usepackage[table]{xcolor}
\usepackage{tikz,tikz-cd}
\usetikzlibrary{shapes.geometric, arrows, positioning}
\usepackage{empheq}
\usepackage{bm}

\pdfoutput=1 
\usepackage{jheppub} 
\usepackage[T1]{fontenc} 

\newcommand{\nn}{\nonumber}

\hypersetup{
    colorlinks=true,
    linkcolor=blue,
    citecolor=blue,
    urlcolor=blue
}

\title{A Tunable Unruh Effect: Accelerated Detectors in Kappa-Rindler Vacua}

\author[a]{Arash Azizi}
\affiliation[a]{{\it The Institute for Quantum Science and Engineering, Texas A\&M University,\\ College Station, TX 77843, U.S.A.}}
\emailAdd{sazizi@tamu.edu}
\date{\today}

\abstract{
We study the response of an accelerated Unruh-DeWitt detector to a one-parameter family of ``kappa Rindler'' vacua, which generalize the standard Unruh effect. These states, parameterized by $\kappa$, continuously interpolate between the Rindler ($\kappa \to 0$) and Minkowski ($\kappa=1$) vacua. We find the detector registers a perfect thermal bath at a tunable temperature $T_\kappa = \kappa T_U$. This result establishes a framework for environments perceived as both ``hotter'' ($\kappa>1$) and ``colder'' ($\kappa<1$) than the standard Unruh temperature. We establish this thermality by demonstrating the KMS condition for the Wightman function and by analyzing the associated particle creation process. Furthermore, we visualize the spacetime structure of the created field quanta, revealing an intuitive link between the $\kappa$-controlled symmetry of the modes and the perceived temperature. Our work provides a comprehensive framework for a modulated Unruh effect, bridging formal QFT with clear visual intuition.
}

\begin{document}
\maketitle
\flushbottom

\section{Introduction}
\label{sec:introduction}

The Unruh effect stands as a cornerstone of modern theoretical physics, predicting that a uniformly accelerated observer will perceive the Minkowski vacuum not as empty space, but as a thermal bath \cite{Unruh1976, Fulling1973, Davies1975}. The temperature of this bath, $T_U = a / (2\pi)$, is directly proportional to the observer's proper acceleration $a$ (in natural units, $\hbar=c=k_B=1$). This profound discovery forged a direct link between acceleration, thermodynamics, and the quantum vacuum, providing crucial insights into Hawking radiation from black holes \cite{Hawking1974, Hawking1975} and shaping our understanding of the observer-dependent nature of particles in quantum field theory \cite{Birrell_Davies1982, Wald1994}.

The Unruh–DeWitt (UDW) \cite{Unruh1976, Einstein100} detector—an idealized two-level system locally coupled to a field—provides a flexible probe of these effects and is widely used in relativistic quantum information (RQI). Reviews map the landscape in which acceleration and horizons reshape quantum resources \cite{Peres2004RMP, Mann2012RQI}, while studies quantify how noninertial motion redistributes and degrades correlations \cite{Alsing2012, Landulfo2009, Martin-Martinez_Leon2010RQI, Martin-Martinez_Fuentes2011RQI, Bruschi2012RQI}. Beyond characterization, operational protocols have been developed in relativistic settings, including communication in the presence of horizons and teleportation using Rindler-mode entanglement \cite{Su2014communication, Downes2013, Foo2020teleportation, Tjoa2022teleportation}; related foundational results highlight information transfer constraints and detector-based measurement frameworks in QFT \cite{Jonsson2015, Polo-Gomez2022measurement}. These works collectively motivate a controllable way to \emph{dial} the effective Unruh temperature while keeping mode structure explicit.

Most analyses of the Unruh effect presuppose that an accelerated detector interacts with the standard Minkowski vacuum. This work, however, explores how the detector's response changes when the surrounding field is prepared in one of the more general ``kappa vacua.'' These states form two distinct families: the boost-stationary \textit{kappa Rindler} vacua \cite{Azizi2022kappashort, Azizi2023JHEP} and the translation-invariant \textit{kappa plane-wave} vacua \cite{Azizi2025KappaPW, Azizi2025KappaGamma}. In this paper, we focus exclusively on the kappa Rindler family, which constitutes a continuous set of physically distinct quantum vacua parameterized by a dimensionless factor $\kappa \ge 0$.

This framework is particularly powerful because it unifies a spectrum of important physical states. The kappa Rindler vacuum contains two celebrated states as special cases:
\begin{itemize}
    \item At $\kappa=1$, it is precisely the Minkowski vacuum, in which an accelerated observer measures the thermal Unruh effect.
    \item As $\kappa \to 0$, the state approaches the Rindler vacuum—the ground state with respect to Rindler time in a single wedge. This limit requires care at the horizon.
\end{itemize}
By treating $\kappa$ as a free parameter, we can move beyond these two specific examples to explore the entire landscape of intermediate quantum vacua. This provides a detailed picture of how a detector's perception of particles is fundamentally shaped by the subtle but crucial properties of the quantum vacuum itself.

We demonstrate that an accelerated detector in a kappa Rindler vacuum registers a precisely thermal response, but with a temperature that is scaled by the parameter $\kappa$, such that $T_\kappa = \kappa T_U$. We establish this result rigorously by showing that the vacuum's two-point Wightman function satisfies the Kubo-Martin-Schwinger (KMS) condition at this modified temperature \cite{Kubo1957, Martin_Schwinger1959}. Complementing this formal proof, we provide an intuitive quantum optics perspective by analyzing the particle creation process responsible for the detector's excitation. We visualize the spacetime structure of the created field quanta—the ``kappa photons''—showing how the parameter $\kappa$ directly controls their symmetry across the light cone. This reinforces the foundational Unruh-Wald insight into the non-local character of the effect \cite{UnruhWald1984} and provides a clear, visual link between the abstract properties of the vacuum state and the concrete thermal response of the detector.

The paper is organized as follows. We begin in Section~\ref{sec:derivation} by deriving the detector's excitation rate from a particle-creation standpoint using first-order perturbation theory. In Section~\ref{sec:Bogol}, we further explore the field's quantum structure by presenting the Bogoliubov transformations that connect different kappa vacua, allowing us to quantify their relative particle content. Section~\ref{sec:wightman} provides a complementary and more formal proof of the thermal response by deriving the Wightman function for the kappa vacuum and verifying that it satisfies the Kubo-Martin-Schwinger (KMS) condition at the modified temperature. In Section~\ref{sec:discussion}, we synthesize these results, discussing the physical implications of a tunable Unruh temperature and using detailed numerical visualizations to provide an intuitive picture of the underlying field modes. Finally, Section~\ref{sec:conclusion} summarizes our findings and outlines potential future directions.

\section{Derivation of the Excitation Rate}
\label{sec:derivation}

\subsection{\texorpdfstring{Notation and setup}{}} \label{subsec:notation}
We work in $(1{+}1)$ dimensions with a massless scalar field and natural units $\hbar=c=k_B=1$. Light–cone coordinates are $u=t-x$ and $v=t+x$. The Heisenberg field is written without a hat for readability,
\[
\Phi(u,v)=\Phi_{\mathrm{RTW}}(u)+\Phi_{\mathrm{LTW}}(v),
\]
while \emph{operator objects} carry hats (creation/annihilation operators, detector operators,  Hamiltonians), e.g.\ $\hat{\mathcal B}_{\Omega,\kappa}$, $\hat{\tilde{\mathcal B}}_{\Omega,\kappa}$, $\hat a_\nu$, $\hat\sigma$, $\hat H_{\text{int}}$. We use $\theta$ for the Heaviside function and $\epsilon\!\to\!0^+$ for the $i\epsilon$ prescription. Uniform acceleration $a$ in the right Rindler wedge corresponds to
\[
-au=e^{-a\tau},\qquad av=e^{a\tau},
\]
with $\tau$ the detector proper time. An IR scale $\mu$ renders logarithms dimensionless.

\subsection{\texorpdfstring{Definition and construction of the $\kappa$–vacua}{}}
\label{subsec:def-kappa}
The $\kappa$–vacua form a one–parameter family of stationary states for each chiral sector (right/left movers). The scalar field $\Phi$ can be decomposed into Right-Traveling Wave (RTW) and Left-Traveling Wave (LTW) components. The chiral field expansions in the $\kappa$–Rindler basis (boost/dilation eigenmodes on the null line; principal branch understood for $(-u)^\kappa$, $u<0$ in the right wedge) are
\begin{align}
\Phi_{\text{RTW}}(u) &= \int_{-\infty}^{+\infty} 
\frac{d\Omega}{\sqrt{\mathcal N_{\Omega,\kappa}}} \bigg(
\theta(-u) (-u)^{i\Omega}e^{\frac{\pi\Omega}{2\kappa}}\hat{\mathcal{B}}_{\Omega,\kappa}
+ \theta(u) u^{i\Omega}e^{-\frac{\pi\Omega}{2\kappa}}\hat{\mathcal{B}}_{\Omega,\kappa}
+\text{h.c.} \bigg), \label{eq:RTW-expand}\\
\Phi_{\text{LTW}}(v) &= \int_{-\infty}^{+\infty} 
\frac{d\Omega}{\sqrt{\mathcal N_{\Omega,\kappa}}} \bigg(
\theta(-v) (-v)^{i\Omega}e^{\frac{\pi\Omega}{2\kappa}}\hat{\tilde{\mathcal B}}_{\Omega,\kappa} 
+ \theta(v) v^{i\Omega}e^{-\frac{\pi\Omega}{2\kappa}}\hat{\tilde{\mathcal B}}_{\Omega,\kappa}
+\text{h.c.} \bigg), \label{eq:LTW-expand}
\end{align}
with the canonical commutators
\[
[\hat{\mathcal B}_{\Omega,\kappa},\hat{\mathcal B}_{\Omega',\kappa}^\dagger]=\delta(\Omega-\Omega'),
\qquad
[\hat{\tilde{\mathcal B}}_{\Omega,\kappa},\hat{\tilde{\mathcal B}}_{\Omega',\kappa}^\dagger]=\delta(\Omega-\Omega'),
\]
and the normalization \(\mathcal N_{\Omega,\kappa}=8\pi\,|\Omega|\,\sinh\!\big(\pi|\Omega|/\kappa\big)\).
The $\kappa$–vacuum is specified by the annihilation conditions
\[
\hat{\mathcal B}_{\Omega,\kappa}\ket{0_\kappa}=0,\qquad
\hat{\tilde{\mathcal B}}_{\Omega,\kappa}\ket{0_\kappa}=0,\qquad \forall\,\Omega\in\mathbb R.
\]
This fixes the state up to an overall phase and is equivalent (see below) to a simple mapping of two–point functions.

This construction is ``Rindler'' because it is built from Rindler modes. The following are the limits of this family:
\begin{itemize}
\item \textbf{Minkowski limit (\(\kappa=1\)).} The chiral maps in \eqref{eq:map-RTW} reduce to the identity, so \(W_\kappa\to W_1\) and the state is exactly the Minkowski vacuum. Uniform acceleration in this vacuum then gives the standard Unruh temperature \(T_U=a/(2\pi)\). The mode content in this limit reduces to the standard Unruh modes \cite{Unruh1976, UnruhWald1984}.
\item \textbf{Rindler limit (\(\kappa\to0^+\)).} The two–point function tends to the ground state with respect to Rindler time in a single wedge (the “Rindler vacuum”). This limit is subtle at the horizon and requires the usual IR care (the $\ln$ and branch choices in \eqref{eq:Wkappa-final-expanded} must be treated before sending \(\kappa\to0\)), but physically the detector temperature \(T_\kappa=\kappa T_U\) tends to zero and excitations vanish.
\item \textbf{Continuum of distinct vacua (\(0<\kappa<\infty\)).} For any other \(\kappa\) one obtains a stationary vacuum that is \emph{neither} Minkowski nor Rindler, yet remains boost–stationary; an accelerated detector measures a strictly thermal bath at \(T_\kappa\). These vacua are mutually Bogoliubov–related via the transformation in Sec.~\ref{sec:Bogol}.
\end{itemize}

This construction and its properties (mode basis, mapping, and limits) were introduced and analyzed in detail in our earlier work~\cite{Azizi2023JHEP}. Here we include the essentials so that the present paper is self-contained; we rely only on \eqref{eq:RTW-expand}–\eqref{eq:LTW-expand} together with the mapping identity \eqref{eq:map-RTW} and the explicit $W_\kappa$ in \eqref{eq:Wkappa-final-expanded}, plus the Bogoliubov analysis in Sec.~\ref{sec:Bogol}.

\subsection{The Interaction Hamiltonian}
The UDW detector is a two-level system with ground state $\ket{g}$ and excited state $\ket{e}$, separated by an energy gap $\omega$. Its interaction with a massless scalar field $\Phi$ is described by the Hamiltonian \cite{UnruhWald1984, Louko2008}:
\begin{equation}
\hat{H}_{\text{int}}(\tau) = g \frac{\partial}{\partial \tau} \Phi(\tau) \left( \hat{\sigma}^\dagger e^{i \omega \tau} + \hat{\sigma} e^{-i \omega \tau} \right),
\label{eq:Hint}
\end{equation}
where $g$ is the coupling constant, $\hat{\sigma}^\dagger = \ket{e}\bra{g}$, and $\hat{\sigma} = \ket{g}\bra{e}$.

The final state $|\Psi_f\rangle$ after the interaction is given by
\begin{equation}
    |\Psi_f^{(1)}\rangle = -\frac{i}{\hbar} \int_{-\infty}^{+\infty} d\tau \hat{H}_{\text{int}}(\tau) |\Psi_i\rangle.
\end{equation}
To study the excitation of the detector, we are interested in the component of the Hamiltonian \eqref{eq:Hint} proportional to the raising operator $\hat{\sigma}^\dagger = |e\rangle\langle g|$, which takes the detector from the ground state to the excited state.

Here, we consider the detector in the \textit{right Rindler wedge}, hence, the relevant components of the field in that region read:
\begin{align}
\Phi^-_{\text{RTW}}(u) &= \int_{-\infty}^{+\infty} 
\frac{d\Omega}{\sqrt{\mathcal{N}_{\Omega,\kappa}}}
(-u)^{-i\Omega}e^{\frac{\pi\Omega}{2\kappa}}\hat{\mathcal B}_{\Omega, \kappa}^\dagger , \nn\\
\Phi^-_{\text{LTW}}(v) &= \int_{-\infty}^{+\infty}
\frac{d\Omega}{\sqrt{\mathcal{N}_{\Omega,\kappa}}}
v^{-i\Omega}e^{-\frac{\pi\Omega}{2\kappa}}\hat{\tilde{\mathcal B}}_{\Omega, \kappa}^{\dagger},
\end{align}
where the superscript ``-'' denotes the negative mode contribution.

For a detector on an accelerated trajectory in the right wedge, its proper time $\tau$ is related to the light-cone coordinates by $-au=e^{-a\tau}$ and $av=e^{a\tau}$. Then we have
\begin{align}
|\Psi_{f, \text{ RTW}}^{(1)}\rangle
&= -\frac{i g}{\hbar} \int_{-\infty}^{+\infty} d\tau \int_{-\infty}^{+\infty} \frac{d\Omega}{\sqrt{\mathcal{N}_{\Omega,\kappa}}} \, \frac{\partial}{\partial \tau}
\left[ \frac{e^{-a\tau}}{a} \right]^{-i\Omega}\, e^{\frac{\pi \Omega}{2\kappa}} e^{i\omega \tau} \, \hat{\mathcal B}_{\Omega, \kappa}^\dagger \ket{0_\kappa}\ket{e}
\nn\\
&= \frac{g}{\hbar} \int_{-\infty}^{+\infty} d\tau \int_{-\infty}^{+\infty} \frac{d\Omega}{\sqrt{\mathcal{N}_{\Omega,\kappa}}} \, a^{1+i\Omega} \Omega\, e^{i(a\Omega + \omega)\tau} e^{\frac{\pi \Omega}{2\kappa}} \,
\hat{\mathcal B}_{\Omega, \kappa}^\dagger \ket{0_\kappa}\ket{e}.
\end{align}
The integral over proper time yields the crucial energy-conservation delta function:
\begin{equation}
    \int_{-\infty}^{+\infty} d\tau \, e^{i(\omega+a\Omega)\tau} = \frac{2\pi}{a} \delta\left(\Omega + \frac{\omega}{a}\right).
\end{equation}
This delta function forces the Rindler frequency of the created quantum to be $\Omega = -\omega/a$. This is a key feature of the Unruh effect: the excitation of the detector (energy gain $\omega$) is accompanied by the creation of a field quantum with negative Rindler frequency.

The final state after the interaction, projected onto the detector's excited state, is found to be
\begin{equation}
|\Psi_{f, \text{ RTW}}^{(1)}\rangle = I(\omega, a, \kappa) |1_{-\frac{\omega}{a}} \rangle_{\kappa} \otimes |e\rangle,
\end{equation}
where $|1_{-\frac{\omega}{a}} \rangle_{\kappa} =\hat{\mathcal{B}}_{-\frac{\omega}{a}, \kappa}^\dagger |0_\kappa\rangle$ is the state of a single created kappa photon and $I(\omega, a, \kappa)$ is its complex amplitude. The probability of this transition is proportional to the modulus squared of this amplitude, $|I(\omega, a, \kappa)|^2$. For a derivative-coupled detector, the calculation yields
\begin{equation*}
     I(\omega,a,\kappa) = \frac{-2\pi g \omega}{\hbar a^{1+i\omega/a}} \frac{e^{-\frac{\pi\omega}{2\kappa a}}}{\sqrt{8\pi\frac{\omega}{a}\sinh\left(\frac{\pi\omega}{\kappa a}\right)}},
\end{equation*}
and hence,
\begin{equation*}
    |I(\omega, a, \kappa)|^2 = \frac{\pi g^2 \omega}{\hbar^2 a} \frac{1}{e^{\frac{2\pi\omega}{\kappa a}}-1}.
\end{equation*}
The transition rate $\mathcal{R}$ is obtained by dividing by the (infinite) interaction time and properly accounting for phase space factors. The final result for the rate contains the characteristic Planck factor for a thermal distribution:
\begin{equation}
\mathcal{R}_{g \to e} \propto \frac{1}{e^{\omega/T_\kappa} - 1},
\end{equation}
where we can read off the perceived temperature:
\begin{equation}
T_\kappa = \frac{\kappa a}{2\pi}.
\end{equation}
This result, derived from the particle creation picture, perfectly matches the temperature found from the KMS condition and the response function method, providing a complete and consistent physical picture.

\section{Particle Content and Bogoliubov Transformations} 
\label{sec:Bogol}

In this section, we analyze the relationship between different kappa vacua by calculating the particle content of one vacuum state as perceived by an observer adapted to another. This is achieved by using the Bogoliubov transformation that connects the creation and annihilation operators of two different kappa-vacua, characterized by parameters $\kappa$ and $\kappa'$.

We begin with the Bogoliubov transformation relating the operators:
\begin{equation}
\hat{\mathcal B}_{\Omega, \kappa'} = \frac{\text{sgn}(\Omega)}{\sqrt{\sinh\left(\frac{\pi\Omega}{\kappa}\right) 
\sinh\left(\frac{\pi\Omega}{\kappa'}\right)}}
\left[ \sinh\left(\frac{\pi\Omega}{2}
\left(\frac{1}{\kappa'} + \frac{1}{\kappa}\right)\right) 
\hat{\mathcal B}_{\Omega, \kappa} 
+ \sinh\left(\frac{\pi\Omega}{2}
\left(\frac{1}{\kappa'} - \frac{1}{\kappa}\right)\right)
\hat{\mathcal B}_{-\Omega, \kappa}^\dagger \right].
\end{equation}
For clarity, we can write this transformation in the standard form $\hat{\mathcal B}_{\Omega, \kappa'} = \alpha_{\Omega,\, \kappa \kappa'} \hat{\mathcal B}_{\Omega, \kappa} + \beta_{\Omega,\, \kappa \kappa'} \hat{\mathcal B}_{-\Omega, \kappa}^{\dagger}$. The number of $\kappa'$-particles of frequency $\Omega$ present in the $\kappa$-vacuum is given by the expectation value:
\begin{equation}
N_{\kappa, \kappa'} (\Omega) = \langle 0_\kappa | \hat{\mathcal B}_{\Omega, \kappa'}^\dagger \hat{\mathcal B}_{\Omega, \kappa'} | 0_\kappa \rangle.
\end{equation}
Substituting the transformation yields:
\begin{equation}
N_{\kappa, \kappa'} (\Omega) = \langle 0_\kappa | (\alpha_{\Omega,\, \kappa \kappa'}^* \hat{\mathcal B}_{\Omega, \kappa}^\dagger + \beta_{\Omega,\, \kappa \kappa'}^* \hat{\mathcal B}_{-\Omega, \kappa}) (\alpha_{\Omega,\, \kappa \kappa'} \hat{\mathcal B}_{\Omega, \kappa} + \beta_{\Omega,\, \kappa \kappa'} \hat{\mathcal B}_{-\Omega, \kappa}^\dagger) | 0_\kappa \rangle.
\end{equation}
Since the annihilation operator $\hat{\mathcal B}_{\Omega, \kappa}$ acting on the vacuum state gives zero, i.e., $\hat{\mathcal B}_{\Omega, \kappa} | 0_\kappa \rangle = 0$ and its conjugate $\langle 0_\kappa | \hat{\mathcal B}_{\Omega, \kappa}^\dagger = 0$, the only non-vanishing term is the one proportional to $|\beta_\Omega|^2$:
\begin{equation}
N_{\kappa, \kappa'} (\Omega) = \langle 0_\kappa | \beta_{\Omega,\, \kappa \kappa'}^* \hat{\mathcal B}_{-\Omega, \kappa} \beta_{\Omega,\, \kappa \kappa'} \hat{\mathcal B}_{-\Omega, \kappa}^\dagger | 0_\kappa \rangle = |\beta_{\Omega,\, \kappa \kappa'}|^2 \langle 0_\kappa | [\hat{\mathcal B}_{-\Omega, \kappa}, \hat{\mathcal B}_{-\Omega, \kappa}^\dagger] | 0_\kappa \rangle.
\end{equation}
The commutator for fields in an infinite volume results in a Dirac delta function, $[\hat{\mathcal B}_{\Omega}, \hat{\mathcal B}_{\Omega'}^\dagger] = \delta(\Omega-\Omega')$, leading to a $\delta(0)$ divergence. This term is understood as being proportional to the total volume of space. The physically meaningful quantity is the particle number per mode (or density), which is obtained by regularizing this divergence. Thus, the particle spectrum is proportional to the Bogoliubov coefficient $|\beta_\Omega|^2$:
\begin{equation}
\boxed{\quad \langle N_{\kappa, \kappa'} (\Omega) \rangle \propto |\beta_{\Omega,\, \kappa \kappa'}|^2 = \frac{\sinh^2\left(\frac{\pi\Omega}{2}\left(\frac{1}{\kappa} - \frac{1}{\kappa'}\right)\right)}{\sinh\left(\frac{\pi\Omega}{\kappa}\right) \sinh\left(\frac{\pi\Omega}{\kappa'}\right)}. \quad}
\end{equation}
As expected, if the two vacua are the same ($\kappa=\kappa'$), the particle number is zero.

\subsection{The Rindler Limit: Recovering the Unruh Effect}
We can now recover the celebrated result for the Unruh effect by considering the particle content of the Minkowski vacuum ($\kappa=1$) as seen from the perspective of a Rindler observer. This corresponds to taking the limit where the Rindler operators are defined by $\kappa' \to 0$.

The number of Rindler particles is proportional to the squared Bogoliubov coefficient $|\beta|^2$. Since this quantity must be positive, and the original expression is even in $\Omega$, we can write it in terms of $|\Omega|$ to ensure the derivation remains straightforward:
$$ N \propto |\beta|^2 = \frac{\sinh^2\left(\frac{\pi|\Omega|}{2}\left(\frac{1}{\kappa'} - 1\right)\right)}{\sinh(\pi|\Omega|) \sinh\left(\frac{\pi|\Omega|}{\kappa'}\right)}$$
In the limit $\kappa' \to 0$, the arguments of the hyperbolic sine functions containing $1/\kappa'$ become large and positive. We can therefore use the approximation $\sinh(x) \approx \frac{1}{2}e^x$ for large $x$. This gives:
\begin{align}
\langle N_{1, 0} (\Omega) \rangle \propto \frac{\left[ \frac{1}{2} e^{\frac{\pi|\Omega|}{2}(\frac{1}{\kappa'}-1)} \right]^2}{\sinh(\pi|\Omega|) \left[ \frac{1}{2} e^{\frac{\pi|\Omega|}{\kappa'}} \right]} = \frac{1}{e^{2\pi|\Omega|}-1}
\end{align}
This result precisely recovers the thermal spectrum for Rindler particles of energy $|\Omega|$ in the Minkowski vacuum. Comparing this to the Planck distribution, $\frac{1}{e^{E/T}-1}$, we see that $2\pi$ plays the role of the inverse Unruh temperature $a/T_U$ (in units where $a=1$), validating the general transformation.

\section{The Wightman Function in \texorpdfstring{$\kappa$}{kappa}-Rindler Spacetime}
\label{sec:wightman}
The detector response is governed by the (unordered) two-point Wightman function
\begin{align}
W(\Delta\tau)=\langle\Phi(\tau)\Phi(\tau')\rangle,
\qquad \Delta\tau\equiv\tau-\tau',
\end{align}
evaluated along the detector worldline. In $(1+1)$ dimensions a massless scalar decomposes into independent right– and left–moving pieces,
$\Phi=\Phi_{\mathrm{RTW}}(u)+\Phi_{\mathrm{LTW}}(v)$ with $u=t-x$ and $v=t+x$. Thus
\begin{align}
W_\kappa=W_{\kappa,\mathrm{RTW}}+W_{\kappa,\mathrm{LTW}}.
\end{align}
We compute $W_{\kappa,\mathrm{RTW}}$ and then add the LTW contribution.

\subsection{\texorpdfstring{Field expansion and $W_{\kappa,\mathrm{RTW}}$}{}}
For the right–moving sector,
\begin{align}
W_{\kappa,\mathrm{RTW}}(u,u')
=\langle 0_\kappa|\Phi_{\mathrm{RTW}}(u)\,\Phi_{\mathrm{RTW}}(u')|0_\kappa\rangle.
\end{align}
Using the $\kappa$–Rindler mode expansion (with $\Omega\in\mathbb{R}$ and $\mathcal N_{\Omega,\kappa}=8\pi|\Omega|\sinh(\pi|\Omega|/\kappa)$),
\begin{align}
\langle 0_\kappa|\Phi(u)\Phi(u')|0_\kappa\rangle
=\!\!\int_{-\infty}^{+\infty}\!\! \frac{d\Omega\,d\Omega'}{\sqrt{\mathcal N_{\Omega,\kappa}\mathcal N_{\Omega',\kappa}}}
\,e^{\frac{\pi\Omega}{2\kappa}}e^{\frac{\pi\Omega'}{2\kappa}}
\langle 0_\kappa|\hat{\mathcal B}_{\Omega,\kappa}\hat{\mathcal B}_{\Omega',\kappa}^\dagger|0_\kappa\rangle
(-u)^{i\Omega}(-u')^{-i\Omega'}.
\end{align}
With $[\hat{\mathcal B}_{\Omega,\kappa},\hat{\mathcal B}_{\Omega',\kappa}^\dagger]=\delta(\Omega-\Omega')$ this reduces to
\begin{align}
W_{\kappa,\mathrm{RTW}}(u,u')
=\int_{-\infty}^{+\infty}\! d\Omega\;
\frac{e^{\frac{\pi\Omega}{\kappa}}}{8\pi\,|\Omega|\,\sinh\!\big(\frac{\pi|\Omega|}{\kappa}\big)}
\left(\frac{-u}{-u'}\right)^{i\Omega}. \label{eq:W-RTW-int}
\end{align}
Rescale $\Omega\mapsto\kappa\Omega$ (with $\kappa>0$) to obtain
\begin{align}
W_{\kappa,\mathrm{RTW}}(u,u')
=\int_{-\infty}^{+\infty}\! d\Omega\;
\frac{e^{\pi\Omega}}{8\pi\,|\Omega|\,\sinh(\pi|\Omega|)}
\left[\left(\frac{-u}{-u'}\right)^{\kappa}\right]^{i\Omega}.
\end{align}
Hence the key mapping property
\begin{align}
\boxed{\;W_{\kappa,\mathrm{RTW}}(u,u')=W_{1,\mathrm{RTW}}((-u)^{\kappa},(-u')^{\kappa})\;}, \label{eq:map-RTW}
\end{align}
i.e. the $\kappa$–RTW Wightman equals the Minkowski–RTW one evaluated on the \emph{flowed} null coordinates.

\subsection*{Calculation in Rindler coordinates}
For a uniformly accelerated worldline in the right wedge with proper acceleration $a$,
$-au=e^{-a\tau}$ and $av=e^{a\tau}$, so
\begin{align}
\frac{u}{u'}=e^{-a\Delta\tau},\qquad \frac{v}{v'}=e^{a\Delta\tau}.
\end{align}
Using the standard Minkowski ($\kappa=1$) RTW/LTW results along this trajectory (see Appendix~\ref{app:Wightman}),
\begin{align}
W_{1,\mathrm{RTW}}(\tau,\tau')&=-\frac{1}{4\pi}\Bigg[\ln\!\frac{2}{a}-\frac{a}{2}(\tau+\tau')+\ln\Big(\sinh\frac{a(\Delta\tau-i\epsilon)}{2}\Big)\Bigg], \label{eq:W-RTW-Mink}\\
W_{1,\mathrm{LTW}}(\tau,\tau')&=-\frac{1}{4\pi}\Bigg[\ln\!\frac{2}{a}+\frac{a}{2}(\tau+\tau')+\ln\Big(\sinh\frac{a(\Delta\tau-i\epsilon)}{2}\Big)\Bigg]. \label{eq:W-LTW-Mink}
\end{align}
The opposite signs of the linear terms reflect $u\propto e^{-a\tau}$ vs.\ $v\propto e^{a\tau}$. Summing RTW+LTW cancels the $\pm\,\frac{a}{2}(\tau+\tau')$, as required by boost stationarity.

By \eqref{eq:map-RTW}, the Wightman function in the $\kappa$–vacuum is obtained by applying the mapping to the standard Minkowski results. This is equivalent to making the substitution $a\to a\kappa$ in Eqs.~\eqref{eq:W-RTW-Mink} and \eqref{eq:W-LTW-Mink}. Let us perform this substitution and sum the two sectors explicitly.

First, we write the $\kappa$-vacuum expressions for the right- and left-moving sectors:
\begin{align}
W_{\kappa,\mathrm{RTW}} &= -\frac{1}{4\pi}\Bigg[\ln\!\frac{2}{a\kappa}-\frac{a\kappa}{2}(\tau+\tau')+\ln\Big(\sinh\frac{a\kappa(\Delta\tau-i\epsilon)}{2}\Big)\Bigg], \\
W_{\kappa,\mathrm{LTW}} &= -\frac{1}{4\pi}\Bigg[\ln\!\frac{2}{a\kappa}+\frac{a\kappa}{2}(\tau+\tau')+\ln\Big(\sinh\frac{a\kappa(\Delta\tau-i\epsilon)}{2}\Big)\Bigg].
\end{align}
Adding these two components to find the total Wightman function $W_\kappa = W_{\kappa,\mathrm{RTW}} + W_{\kappa,\mathrm{LTW}}$, the linear terms in $(\tau+\tau')$ cancel, as required by stationarity:
\begin{align}
W_\kappa(\Delta\tau) &= -\frac{1}{4\pi} \left[ 2\ln\frac{2}{a\kappa} + 2\ln\Big(\sinh\frac{a\kappa(\Delta\tau-i\epsilon)}{2}\Big) \right] \nonumber \\
&= -\frac{1}{4\pi} \ln\left[ \frac{4}{a^2\kappa^2} \sinh^2\Big(\frac{a\kappa(\Delta\tau-i\epsilon)}{2}\Big) \right].
\end{align}
Introducing an IR scale $\mu$ to render the logarithm dimensionless, the total unordered Wightman function is
\begin{align}
\boxed{\; 
W_\kappa(\Delta\tau)
=\langle 0_\kappa|\Phi(\tau)\Phi(\tau')|0_\kappa\rangle
= -\frac{1}{4\pi}\,
\ln\!\left[
\mu^2\,\frac{4}{a^2\kappa^2}\,
\sinh^2\!\Big(\frac{a\kappa}{2}\,(\Delta\tau-i\epsilon)\Big)
\right]. \;
} \label{eq:Wkappa-final-expanded}
\end{align}
\vspace{.5cm}

\subsection{KMS condition}
The thermal property of the $\kappa$-vacuum is encoded by the KMS (Kubo-Martin-Schwinger) condition, which relates the greater/lesser Wightman functions  $W^>(\tau) = \langle \Phi(\tau)\Phi(0) \rangle$ and $W^<(\tau) = \langle \Phi(0)\Phi(\tau) \rangle$ at real and imaginary times:
\begin{align}
\boxed{\; W^>(\tau-i\beta_\kappa) = W^<(\tau) , \;}
\qquad \beta_\kappa=\frac{2\pi}{a\kappa}.
\end{align}
For a free scalar field, these ordered functions are boundary values of the unordered Wightman function, $W^>(\tau) = W_\kappa(\tau-i\epsilon)$ and $W^<(\tau) = W_\kappa(\tau+i\epsilon)$. We now verify this relation using the explicit form of $W_\kappa(\Delta\tau)$ from Eq.~\eqref{eq:Wkappa-final-expanded}.

\paragraph{Verification.} We evaluate the two sides of the KMS relation separately.
The Left-Hand Side (LHS) is
\begin{align}
\mathrm{LHS} &= W^>(\tau-i\beta_\kappa) = W_\kappa(\tau-i\beta_\kappa - i\epsilon) \nonumber\\
&= -\frac{1}{4\pi}\ln\!\left[
\mu^2\,\frac{4}{a^2\kappa^2}\,
\sinh^2\!\Big(\frac{a\kappa}{2}(\tau-i\beta_\kappa-i\epsilon)\Big)
\right].
\end{align}
Using the identity $(a\kappa/2)\beta_\kappa=\pi$ and the property $\sinh(z-i\pi)=-\sinh z$, the argument of the logarithm simplifies:
\begin{align}
\sinh^2\!\Big(\frac{a\kappa}{2}(\tau-i\epsilon) - i\pi \Big)
=\sinh^2\!\Big(\frac{a\kappa}{2}(\tau-i\epsilon)\Big).
\end{align}
Thus, the LHS reduces to the boundary value of the Wightman function from below the real axis:
\begin{align}
\mathrm{LHS} = -\frac{1}{4\pi}\ln\!\left[
\mu^2\,\frac{4}{a^2\kappa^2}\,
\sinh^2\!\Big(\frac{a\kappa}{2}(\tau-i\epsilon)\Big)
\right] = W_\kappa(\tau-i\epsilon).
\end{align}
The Right-Hand Side (RHS) is, by definition,
\begin{align}
\mathrm{RHS} = W^<(\tau) = W_\kappa(\tau+i\epsilon).
\end{align}
For the KMS condition to be satisfied, we must have LHS = RHS, which requires $W_\kappa(\tau-i\epsilon)=W_\kappa(\tau+i\epsilon)$. For real $\tau\neq0$, the argument of the logarithm in Eq.~\eqref{eq:Wkappa-final-expanded} is strictly positive, meaning the function is analytic on the real axis away from the origin. Therefore, the boundary values from above and below coincide:
\begin{align}
W_\kappa(\tau-i\epsilon)=W_\kappa(\tau+i\epsilon),
\qquad \tau\neq0.
\end{align}
This confirms that $\mathrm{LHS}=\mathrm{RHS}$, proving the KMS relation is satisfied with the inverse temperature $\beta_\kappa$, which corresponds to the perceived temperature:
\begin{align}
T_\kappa=\beta_\kappa^{-1}=\frac{a\kappa}{2\pi}.
\end{align}
At the coincidence point $\tau=0$, the $i\epsilon$ prescription correctly regulates the singularity of the Wightman function in the distributional sense.

\section{A Spectrum of Thermal Vacua: Visualizing the Unruh Effect}
\label{sec:discussion}
The agreement between our two derivations and the verification of the KMS condition robustly establish our main result: an accelerated detector in a kappa vacuum experiences a perfect thermal bath at temperature $T_\kappa = \kappa T_U$.

\subsection{Analysis of the Thermal Response}
\begin{figure}[ht!]
 \centering
 \includegraphics[width=0.9\textwidth]{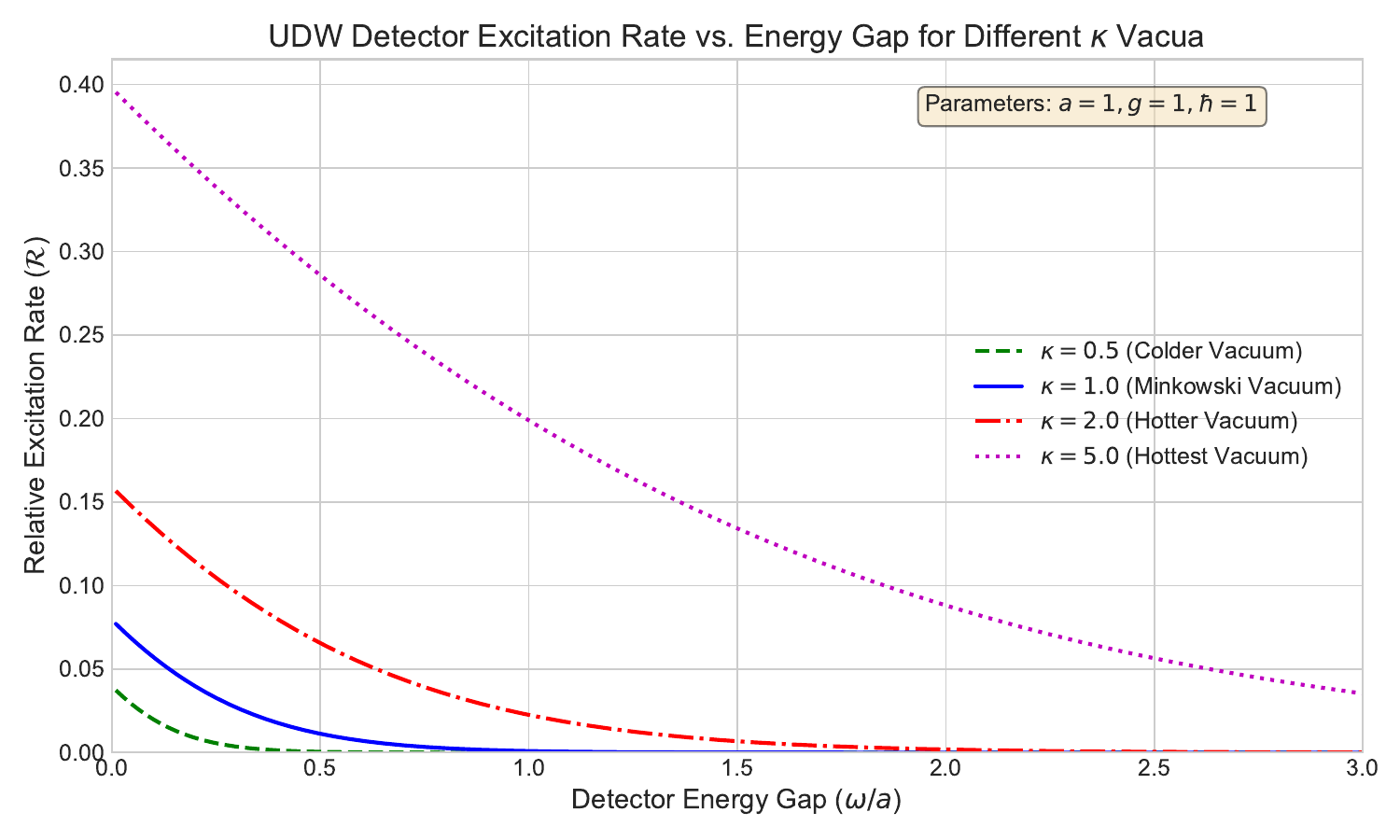}
 \caption{The relative excitation rate $\mathcal{R}$ of a uniformly accelerated Unruh-DeWitt detector as a function of its energy gap $\omega/a$. The curves correspond to different kappa vacua, characterized by the parameter $\kappa$. The case $\kappa=1$ (solid blue) represents the standard Unruh effect. For $\kappa < 1$ (dashed green), the thermal response is suppressed. For $\kappa > 1$ (dot-dashed red and dotted magenta), the perceived temperature and excitation rate are enhanced. Constants are set to unity ($a=1, g=1, \hbar=1$).}
\label{fig:rate_comparison}
\end{figure}

The parameter $\kappa$ acts as a ``dial'' for the perceived vacuum temperature. This is illustrated in Figure~\ref{fig:rate_comparison}, which plots the excitation rate versus the detector's energy gap for several values of $\kappa$. For $\kappa=1.0$ (solid blue), we recover the standard Unruh effect. For $\kappa < 1$, the rate is strongly suppressed, corresponding to a colder vacuum. For $\kappa > 1$ (red and magenta lines), the rate is enhanced, indicating a hotter vacuum. The limits are physically intuitive:
\begin{itemize}
 \item $\kappa \to 1$: The kappa vacuum becomes the Minkowski vacuum.
 \item $\kappa \to 0$: $T_\kappa \to 0$, approaching the Rindler vacuum, i.e. the ground state with respect to Rindler time translations in a wedge \cite{Fulling1973,Unruh1976,Wald1994,Birrell_Davies1982}. The excitation rate vanishes in this limit (with IR care at the horizon).
\end{itemize}

\subsection{The Unruh-Wald Insight: Locus of the Emitted Photon}
\begin{figure}[ht!]
    \centering
    \includegraphics[width=\textwidth]{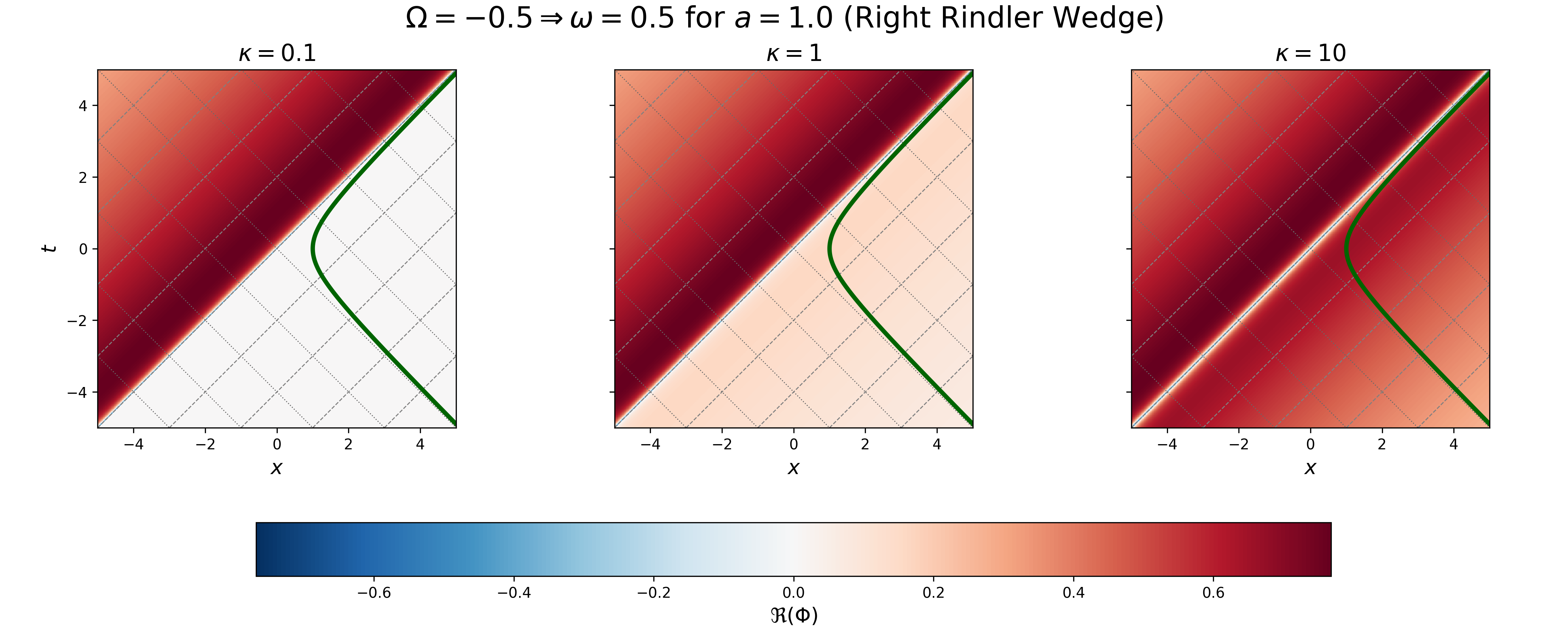}
    \caption{Spacetime structure of a right-traveling kappa mode for a detector energy gap $\omega=0.5a$ (corresponding to Rindler frequency $\Omega=-0.5$). The real part of the mode function $\Re(\Phi)$ is plotted for vacuum parameters $\kappa=0.1$ (left), $\kappa=1.0$ (middle), and $\kappa=10.0$ (right). The green curve is the worldline of an observer with acceleration $a=1$. The dashed grid lines are light-cone coordinates.}
    \label{fig:kappa_mode_0.5}
\end{figure}

\begin{figure}[ht!]
    \centering
    \includegraphics[width=\textwidth]{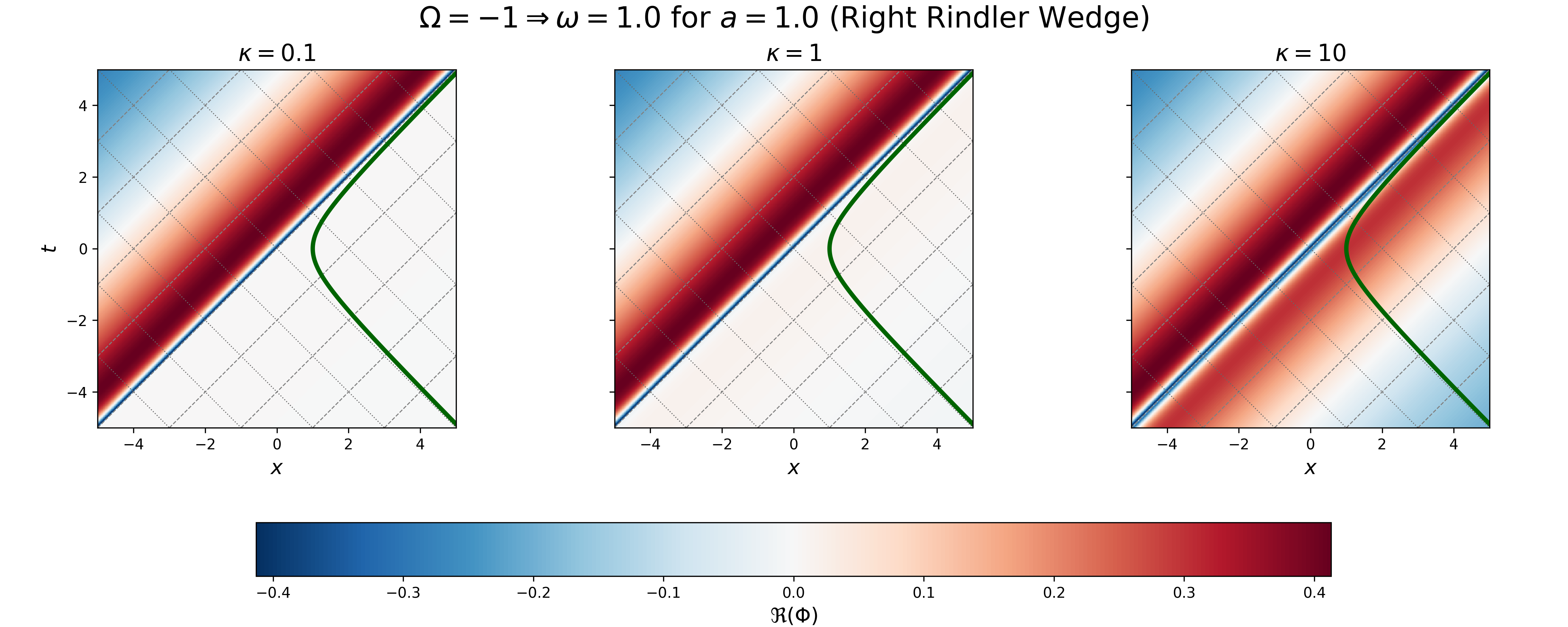}
    \caption{Spacetime structure of a right-traveling kappa mode for a detector energy gap $\omega=1.0a$ ($\Omega=-1.0$). The panels show the mode for $\kappa=0.1$, $\kappa=1.0$, and $\kappa=10.0$. As $\kappa$ increases, the mode becomes more symmetric across the $t=x$ light cone, corresponding to a higher perceived temperature.}
    \label{fig:kappa_mode_1.0}
\end{figure}

\begin{figure}[ht!]
    \centering
    \includegraphics[width=\textwidth]{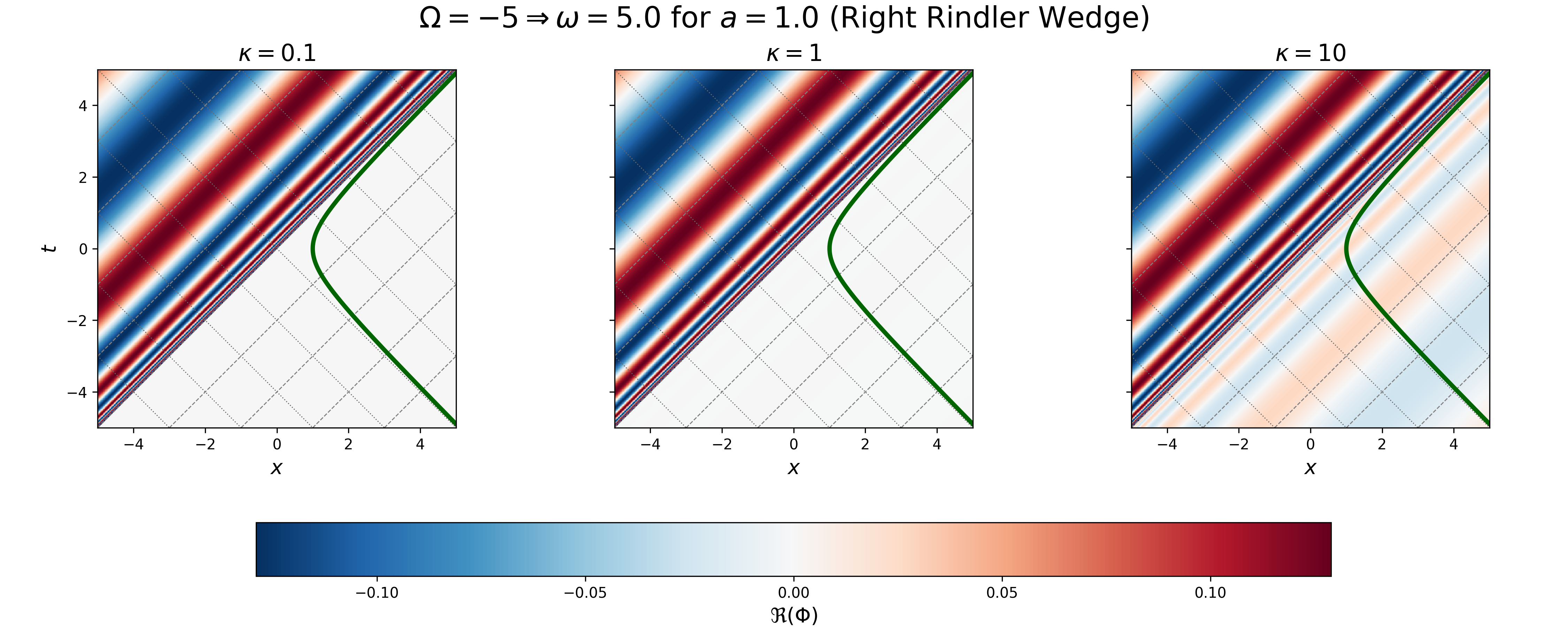}
    \caption{Spacetime structure of a right-traveling kappa mode for a detector energy gap $\omega=5.0a$ ($\Omega=-5.0$). The panels show the mode for $\kappa=0.1$, $\kappa=1.0$, and $\kappa=10.0$. The high frequency of the mode is apparent in the rapid oscillations. The trend of increasing symmetry with increasing $\kappa$ holds, confirming that the thermal properties of the vacuum are independent of the specific mode frequency being probed.}
    \label{fig:kappa_mode_5.0}
\end{figure}

The particle creation perspective reveals a profound, non-local feature of the Unruh effect. The field quantum created when the detector clicks, a ``kappa photon,'' is described by the state $\hat{\mathcal B}^\dagger_{-\omega/a, \kappa} |0_\kappa\rangle$. As discovered by Unruh and Wald, this quantum is not localized with the detector \cite{UnruhWald1984}. The structure of this mode, which dictates the thermal response, is visualized in Figures~\ref{fig:kappa_mode_0.5}, \ref{fig:kappa_mode_1.0}, and \ref{fig:kappa_mode_5.0}. These figures show the real part of the right-traveling kappa mode function for various values of the vacuum parameter $\kappa$ and for three different detector energy gaps, $\omega = -a\Omega$.

The plots clearly illustrate how the parameter $\kappa$ acts as a dial for the thermal properties of the vacuum by controlling the spacetime symmetry of the field modes. Across all three figures, we observe a consistent trend:
\begin{itemize}
    \item For a ``cold'' vacuum ($\kappa=0.1$, left panels in each figure), the mode is almost entirely confined to the region $u=t-x<0$, which contains the detector's trajectory. The amplitude in the causally disconnected region $u>0$ is exponentially suppressed, consistent with the vanishing particle detection rate in the Rindler vacuum limit ($\kappa \to 0$).
    \item For the standard Minkowski vacuum ($\kappa=1.0$, middle panels), we see the mode-mixing characteristic of the Unruh effect. There is now a significant mode amplitude in the $u>0$ region, though it remains subdominant to the amplitude in the detector's wedge.
    \item For a ``hot'' vacuum ($\kappa=10.0$, right panels), the mode becomes nearly symmetric across the light-cone $u=0$. This indicates a strong mixing of positive and negative Rindler frequencies, which the detector perceives as a higher temperature, $T_\kappa = \kappa T_U$.
\end{itemize}

Furthermore, by comparing Figures~\ref{fig:kappa_mode_0.5}, \ref{fig:kappa_mode_1.0}, and \ref{fig:kappa_mode_5.0}, we can see the effect of the detector's energy gap $\omega$. A larger energy gap (and thus larger $|\Omega|$) leads to more rapid spacetime oscillations of the mode. It also leads to a greater degree of asymmetry for any fixed $\kappa$. The ratio of the mode's amplitude in the $u>0$ region to the $u<0$ region is proportional to $e^{-2\pi\omega/\kappa a}$, which decreases for larger $\omega$. This implies that higher-energy detector excitations are sourced by field quanta that are more strongly localized within the detector's own Rindler wedge. The bigger $|\Omega|$, the more \textit{asymmetric} the mode becomes for a fixed $\kappa$.

\section{Conclusion}
\label{sec:conclusion}
We have investigated the response of a uniformly accelerated UDW detector to the family of kappa vacuum states. We have shown that the detector's response is rigorously thermal with a perceived temperature $T_\kappa = \kappa T_U$. This was first established formally by verifying the KMS condition for the vacuum's Wightman function, and then re-derived from a quantum optics perspective that highlights the particle creation process. We emphasized that this process is sourced by negative Rindler frequency modes and results in a non-local ``kappa photon'' predominantly located in the causally disconnected Rindler wedge, a key feature discovered by Unruh and Wald.

This work establishes the kappa vacuum as a versatile theoretical tool, providing a continuous parameter $\kappa$ to smoothly transition between the cold Rindler vacuum ($\kappa=0$), the familiar Minkowski vacuum ($\kappa=1$), and hotter-than-Unruh thermal environments ($\kappa>1$). Our findings generalize the Unruh effect and provide a concrete framework for studying observer-dependent phenomena. Future work could extend this analysis to higher-order processes \cite{Svidzinsky2021PRL, Doukas2009} or investigate RQI tasks \cite{Bruschi2010QI_Unruh, Aspling:2024wbz, Jonsson2015} in this tunable vacuum background.

\begin{acknowledgments}
I am grateful to Girish Agarwal, Marlan Scully, Bill Unruh, and Suhail Zubairy for discussions. This work was supported by the Robert A. Welch Foundation (Grant No. A-1261) and the National Science Foundation (Grant No. PHY-2013771).
\end{acknowledgments}
\appendix

\section{Wightman Function in Minkowski and Rindler Coordinates}
\label{app:Wightman}

\subsection{Minkowski Coordinates}
\label{app:Wightman_mink}

Let us first evaluate the Wightman function for the Minkowski case ($\kappa=1$). While the integral for $W_{1, \text{RTW}}(u,u')$ can be computed directly, it is simpler to use the standard plane-wave expansion for the field in the Minkowski vacuum $|0_M\rangle$:
\begin{align}
W_{1, \text{RTW}}(u,u')&=\langle 0_M | \Phi_{\text{RTW}}(u) \Phi_{\text{RTW}}(u') | 0_M \rangle = 
\int_0^\infty \int_0^\infty d\nu d\nu' \frac{e^{-i\nu u}}{\sqrt{4\pi \nu}} \frac{e^{+i\nu'u'}}{\sqrt{4\pi \nu'}} \underbrace{\langle 0_M | \hat{a}_\nu \hat{a}_{\nu'}^\dagger | 0_M \rangle}_{\delta(\nu-\nu')} \nn \\
&= \frac{1}{4\pi} \int_0^\infty \frac{d\nu}{\nu} e^{-i\nu(u-u')}.
\end{align}
This is a standard integral, see the appendix~\ref{app:int} for detailed derivation, which evaluates to:
\begin{align}
W_{1, \text{RTW}}(u,u') =& -\frac{1}{4\pi} \ln\!\big[\mu\,(u-u'-i\epsilon)\big]
 \nn\\
=& -\frac{1}{4\pi} \Big( \theta(u-u')\ln\!\big[\mu(u-u')\big] + \theta(u'-u)\big(\ln\!\big[\mu(u'-u)\big]-i\pi\big) \Big)
\label{Wightman_Mink}
\end{align}

\subsection{Right-Moving Modes Rindler Coordinates}
\label{app:Wightman_Rind_RTW}
Here we present a detailed calculation of the Wightman function for right-moving modes in terms of Rindler coordinates. The goal is to express the standard Wightman function for a chiral field, given by $W_1 = -\frac{1}{4\pi} \ln(u-u'-i\epsilon)$, in terms of the proper time $\tau$ of an accelerated observer.

For a uniformly accelerated observer with proper acceleration $a$, the light-cone coordinate $u$ is related to their proper time $\tau$ by $u = -(1/a)e^{-a\tau}$. We are interested in the quantity $u-u'$, which is the argument of the logarithm.
\begin{align}
u - u' &= \left(-\frac{1}{a}e^{-a\tau}\right) - \left(-\frac{1}{a}e^{-a\tau'}\right) \nonumber\\
&= \frac{1}{a} e^{-a\frac{\tau+\tau'}{2}} \left(e^{a\frac{\tau-\tau'}{2}} - e^{-a\frac{\tau-\tau'}{2}}\right) \nonumber\\
&= \frac{2}{a} e^{-a\frac{\tau+\tau'}{2}} \sinh\left(\frac{a\Delta\tau}{2}\right),
\end{align}
where we have defined the proper time separation as $\Delta\tau = \tau-\tau'$.

Substituting this result back into the expression for the Wightman function (\ref{Wightman_Mink}), we get:
\begin{align}
W_1(\tau, \tau') = -\frac{1}{4\pi} \ln\left[ \frac{2}{a} e^{-a\frac{\tau+\tau'}{2}} \sinh\left(\frac{a\Delta\tau}{2}\right) - i\epsilon \right].
\end{align}
Using the property of logarithms, $\ln(xyz) = \ln x + \ln y + \ln z$, we can separate the terms. The small imaginary part $-i\epsilon$ regulates the logarithm when its argument is negative or zero.
\begin{align}
W_1(\tau, \tau') = -\frac{1}{4\pi} \left\{ \ln\left(\frac{2}{a}\right) - a\frac{\tau+\tau'}{2} + \ln\left[ \sinh\left(\frac{a\Delta\tau}{2}\right) - i\epsilon' \right] \right\},
\end{align}
where $\epsilon'$ is another infinitesimal positive constant that has absorbed the prefactors. The final logarithmic term is a standard way of representing the complex logarithm with the correct branch choice. It can be written more compactly by placing the $-i\epsilon$ inside the argument of the sinh function, which produces the same result.

This gives the final, correct expression for the right-moving Wightman function:
\begin{equation}
\boxed{\quad
W_{1,\mathrm{RTW}}(\tau, \tau') = -\frac{1}{4\pi} \left[ \ln\frac{2\mu}{a} - \frac{a}{2}(\tau+\tau') + \ln\!\left(\sinh\frac{a(\Delta\tau-i\epsilon)}{2}\right) \right].
\quad}
\end{equation}

\subsection{Left-Moving Modes in Rindler Coordinates}
\label{app:Wightman_Rind_LTW}

Here we repeat the argument for the left-moving modes, described by the light-cone coordinate $v=t+x$. The derivation is analogous to the right-moving case, with the primary difference arising from the transformation law for the $v$ coordinate.

The Wightman function for a left-moving chiral field in the Minkowski vacuum is given by:
\begin{equation}
W(v, v') = \langle 0_M | \Phi(v) \Phi(v') | 0_M \rangle = -\frac{1}{4\pi} \ln(v-v'-i\epsilon).
\end{equation}
For a uniformly accelerated observer with proper acceleration $a$ in the right Rindler wedge, the light-cone coordinate $v$ is related to their proper time $\tau$ by:
\begin{equation}
v = \frac{1}{a} e^{a\tau}.
\end{equation}
We now express the argument of the logarithm, $v-v'$, in terms of the observer's proper time.
\begin{align}
v - v' &= \frac{1}{a}\left(e^{a\tau} - e^{a\tau'}\right) \nonumber\\
&= \frac{1}{a} e^{a\frac{\tau+\tau'}{2}} \left(e^{a\frac{\tau-\tau'}{2}} - e^{-a\frac{\tau-\tau'}{2}}\right) \nonumber\\
&= \frac{2}{a} e^{a\frac{\tau+\tau'}{2}} \sinh\left(\frac{a\Delta\tau}{2}\right).
\end{align}
Substituting this result back into the expression for the Wightman function and defining the proper time separation as $\Delta\tau = \tau-\tau'$, we get:
\begin{align}
W_{1,\mathrm{LTW}}(\tau, \tau') &= -\frac{1}{4\pi} \ln\left[ \frac{2}{a} e^{a\frac{\tau+\tau'}{2}} \sinh\left(\frac{a\Delta\tau}{2}\right) - i\epsilon \right] \nonumber\\
&= -\frac{1}{4\pi} \left\{ \ln\left(\frac{2}{a}\right) + a\frac{\tau+\tau'}{2} + \ln\left[ \sinh\left(\frac{a\Delta\tau}{2}\right) - i\epsilon' \right] \right\}.
\end{align}
Note that the sign of the $\frac{a}{2}(\tau+\tau')$ term is positive, in contrast to the right-moving case, due to the positive exponent in the transformation for $v$.

This gives the final, correct expression for the left-moving Wightman function in Rindler coordinates:
\begin{equation}
\boxed{\quad
W_{1,\mathrm{LTW}}(\tau, \tau') = -\frac{1}{4\pi} \left[ \ln\frac{2\mu}{a} + \frac{a}{2}(\tau+\tau') + \ln\!\left(\sinh\frac{a(\Delta\tau-i\epsilon)}{2}\right) \right].
\quad}
\end{equation}

\section{\texorpdfstring{\boldmath Finding the integral $f(k)$}{}} 
\label{app:int}

We need
\begin{align}
f(k) \equiv \int_{0}^{\infty} \frac{e^{-izk}}{z}\,dz.
\end{align}
We evaluate it in three different ways.

\subsection{Integral evaluation via contour integration}

Consider
\begin{align}
f(k)-f(-k)
&=\int_{0}^{\infty}\frac{e^{-izk}}{z}\,dz-\int_{0}^{\infty}\frac{e^{izk}}{z}\,dz \\
&=\mathrm{P.V.}\!\int_{-\infty}^{\infty}\frac{e^{-izk}}{z}\,dz.
\end{align}
For $k>0$, $e^{-izk}=e^{yk}e^{-ixk}$ decays in the lower half–plane. Closing the contour there with an indentation below the origin gives
\begin{align}
\mathrm{P.V.}\!\int_{-\infty}^{\infty}\frac{e^{-izk}}{z}\,dz
&=-\int_{\pi}^{0} i\,d\theta
=-i\pi .
\end{align}
For $k<0$ the contour is closed above, giving $+i\pi$. Hence
\begin{align}
f(k)-f(-k)=-i\pi\,\mathrm{sgn}\,k. \label{eq:diff_f}
\end{align}

\subsection{Integral evaluation via differentiation}

Differentiate under the integral sign (with regulator $\epsilon>0$):
\begin{align}
f'(k)&=\frac{d}{dk}\int_{0}^{\infty}\frac{e^{-izk}}{z}\,dz
=-i\int_{0}^{\infty}e^{-iz(k-i\epsilon)}dz
=-\frac{1}{k-i\epsilon}.
\end{align}
Integrating back,
\begin{align}
f(k)=-\ln(k-i\epsilon)+C,
\end{align}
with $C$ a constant. Using
\begin{align}
\ln(k-i\epsilon)=\theta(k)\ln k+\theta(-k)\big(\ln|k|-i\pi\big),
\end{align}
we obtain
\begin{align}
f(k)=-\theta(k)\ln k-\theta(-k)\big(\ln|k|-i\pi\big)+C. \label{eq:branch_form}
\end{align}
Eq.~\eqref{eq:branch_form} reproduces \eqref{eq:diff_f} for any $C$.

\subsection{Evaluation via tabulated integrals}

Introduce a real regulator $\epsilon>0$ and split into sine/cosine parts,
\begin{align}
I_\epsilon(k)&=\int_{0}^{\infty}\frac{e^{-(\epsilon+ik)z}}{z}\,dz
=\int_{0}^{\infty}\frac{e^{-\epsilon z}\cos(kz)}{z}\,dz
-i\int_{0}^{\infty}\frac{e^{-\epsilon z}\sin(kz)}{z}\,dz .
\end{align}
From Gradshteyn \& Ryzhik \cite{Gradshteyn_Ryzhik2014}:  
\begin{itemize}
    \item §3.948.2: $\displaystyle \int_{0}^{\infty}e^{-\beta x}\frac{\cos ax-\cos bx}{x}\,dx=\tfrac12\ln\frac{b^2+\beta^2}{a^2+\beta^2}$ ($\Re\beta>0$).  
    \item §3.941.1: $\displaystyle \int_{0}^{\infty}e^{-px}\frac{\sin(qx)}{x}\,dx=\arctan\frac{q}{p}$ ($p>0$). 
    \item §3.951.3: $\displaystyle \int_{0}^{\infty}\frac{e^{-\gamma x}-e^{-\beta x}}{x}\cos(bx)\,dx=\tfrac12\ln\frac{b^2+\beta^2}{b^2+\gamma^2}$ ($\Re\beta>0,\Re\gamma\ge0$).
\end{itemize}
Applying these with $a=k$, $b=0$, $\beta=\epsilon$ gives
\begin{align}
\Re I_\epsilon(k)&=-\tfrac12\ln\!\Big(1+\frac{k^2}{\epsilon^2}\Big)+\text{const},\\
\Im I_\epsilon(k)&=-\arctan\!\frac{k}{\epsilon}.
\end{align}
Letting $\epsilon\to0^+$ and introducing an IR scale $\mu>0$,
\begin{align}
\int_{0}^{\infty}\frac{e^{-ikz}}{z}\,dz
=-\ln\!\frac{|k|}{\mu}-i\,\frac{\pi}{2}\,\mathrm{sgn}\,k. \label{eq:SiCi_form}
\end{align}

\subsection{Consistency and choice of constant}

Equations \eqref{eq:branch_form} and \eqref{eq:SiCi_form} differ by a $k$-independent constant:
\begin{align}
-\ln(k-i\epsilon)
=\Big[-\ln\!\frac{|k|}{\mu}-i\,\frac{\pi}{2}\,\mathrm{sgn}\,k\Big]+\Big(-\ln\mu+\tfrac{i\pi}{2}\Big).
\end{align}
Thus the two approaches are equivalent: the difference is just an additive constant, which can be absorbed into the arbitrary scale $\mu$. Physically, only differences such as $f(k)-f(-k)$ matter, and both methods give the same result.

\bibliographystyle{jhep}
\bibliography{UnruhRef}
\end{document}